# Citation success index – An intuitive pair-wise journal comparison metric

Staša Milojević[1], Filippo Radicchi[1] and Judit Bar-Ilan[2]

[1]School of Informatics and Computing, Indiana University, Bloomington, USA

[2]Department of Information Science, Bar-Ilan University, Ramat Gan, 5290002, Israel

In this paper we present "citation success index", a metric for comparing the citation capacity of pairs of journals. Citation success index is the probability that a random paper in one journal has more citations than a random paper in another journal (50% means the two journals do equally well). Unlike the journal impact factor (IF), the citation success index depends on the broadness and the shape of citation distributions. Also, it is insensitive to sporadic highly-cited papers that skew the IF. Nevertheless, we show, based on 16,000 journals containing ~2.4 million articles, that the citation success index is a relatively tight function of the ratio of IFs of journals being compared, due to the fact that journals with same IF have quite similar citation distributions. The citation success index grows slowly as a function of IF ratio. It is substantial (>90%) only when the ratio of IFs exceeds ~6, whereas a factor of two difference in IF values translates into a modest advantage for the journal with higher IF (index of ~70%). We facilitate the wider adoption of this metric by providing an online calculator that takes as input parameters only the IFs of the pair of journals.

1. **Introduction**

Most authors of research articles, whether in teams or as individuals, ultimately aim to maximize the impact of their publications, even when this goal is expressed as a desire to reach the widest possible audience (Gordon, 1984; Luukkonen, 1992). The simplest and most direct indication of an impact of a publication is the number of citations it has received over some period of time. Despite warnings from the scientometrics community against the inappropriate interpretation of the research metrics (e.g., Hicks, Wouters, Waltman, de Rijcke, & Rafols, 2015), the author's citation count and the related h-index (Hirsch, 2005) can still be critical factors for funding, hiring, tenure and promotion decisions (Wouters, 2014).

To increase the visibility of their work within the scientific community, and eventually help increase their citation counts, authors often engage in various activities, such as presenting their work at conferences, giving colloquia, etc. Many authors believe that having a publication in a higher-impact venue is yet another avenue for increasing the visibility of their work, which may lead to receiving more citations and consequently more rewards (Calcagno et al., 2012). In particular, the authors often aspire to publish in high-impact, general-science journals, rather than the less prestigious specialized venues (Verma, 2015). Even when choosing among alternative specialized venues, authors tend to give preference to higher ranked ones (Garfield, 2006; S. Rousseau & Rousseau, 2012). A recent ethnographic study that examined the role of the performance metrics in knowledge production has found that some researchers think "that articles appearing in high impact journals generally attract larger citation numbers *precisely because they are published in high impact journals*" (Rushforth & De Rijcke, 2015, p. 133). Consequently, some authors adopt a practice of targeting the highest-impact venue first, "cascading" to journals with lower impact until acceptance (Gordon, 1984), even though this process can exert significant publication delays and place a burden on editors and reviewers, as well as the authors. It is outside of the scope of this paper to try to establish to what extent are such attitudes correct. There is some empirical evidence that very similar articles published in journals with higher impact factors do end up receiving more citations than their "twins" published in lower impact factor journals (Larivière & Gingras, 2010; Perneger, 2010; Shanahan, 2016). The prestige of a journal is often used, implicitly if not explicitly, as an assessment of the quality of research (De Rijcke, Wouters, Rushforth, Franssen, & Hammarfelt, 2016; Ravetz, 1971), so it is not surprising that the papers published in more prestigious journals will reach a wider audience, especially considering that many researchers do not have the time to learn about all the research being published in their research areas, instead giving priority to higher-ranked journals in their field and to prestigious general-science journals. (De Rijcke et al., 2016; Rushforth & De Rijcke, 2015).

Journal impact factor (IF), often considered "a direct reflection of a journal's prestige or quality" (Moed, 2010, p. 91), is the most widely used journal impact measure (Glänzel & Moed, 2002). The IF is a metric introduced by Eugene Garfield in 1972 (Garfield, 1972), and its definition is rather simple. The IF of a venue in year $y$ equals the number of citations received in $y$ to all documents published in that venue in the preceding two years ($y - 2$ and $y - 1$), divided by the number of "citable documents" (defined as research articles and reviews) covered by the citation database (Moed & van Leeuwen, 1996). Official IF values are released annually by the Thomson Reuters Journal Citation Reports.

Despite the prevalence of an IF as a measure of journal impact, there is a large body of research arguing that evaluating the impact of journals is not a straightforward task (Bar-Ilan, 2012; Bornmann, Werner, Gasparyan, & Kitas, 2012; Haustein, 2012; R. Rousseau, 2002; Thelwall, 2012; Waltman, 2016). For example, the database coverage has a strong effect on the IF, thus disadvantaging fields with strong non-English literature (e.g., social sciences and humanities) (Leydesdorff & Milojević, 2015) or the ones that publish heavily in non JCR-indexed literature (e.g., computer science and humanities) (Althouse, West, Bergstrom, & Bergstrom, 2009). Nevertheless, IF has been used by journal editors and publishers to attract submissions and readership, and by researchers as an indicator of prestige and as a tool for screening an ever-growing body of literature for reading and, eventually, citing (De Rijcke et al., 2016; Rushforth & De Rijcke, 2015).

The most contested and criticized usage of IFs has been to assess, at least in the short-term, the quality of individual scientific publications on the basis of the IF of the venue (e.g., Archambault & Larivière, 2009; DORA, 2012). At the heart of this criticism lies the fact that the IF is a very poor predictor of the number of citations that a given paper will receive (Seglen, 1992, 1997). The reasons for this are essentially two-fold: (1) the citation distributions for individual journals are broad and therefore overlap even if their IFs are quite different (Larivière et al., 2016). (2) Furthermore, these wide citation distributions are skewed, so that the IF, being based on an arithmetic mean, may be affected by the tail of a small number of highly-cited articles. These limitations of the IF have led to a recent proposal that journals should advertise full citation distributions (the number (or fraction) of papers that have received 0, 1, 2,… citations) rather than just the IFs (Larivière et al., 2016).

In this paper, we present a new metric that is specifically designed to compare pairs of journals and addresses the aforementioned limitations of the IF. The metric, which we call *citation success index*, depends on the shape (e.g., the broadness) of the citation distributions. Also, unlike the IF it is not skewed by sporadic highly-cited papers. We define the citation success index as the probability that a random paper in journal A has more citations than a random paper in journal B. This metric not only acknowledges the fact that some articles from a low-IF journal may receive more citations than some articles from a higher-IF journal, a point that was made in Larivière et al. (2016), but actually quantifies the likelihood of such outcome in a simple and intuitive way.

2. **Materials and methods**

For this study we use *Thomson Reuters* Web of Science (WoS) database of bibliographic records of journal articles. Specifically, we use all records that WoS classifies as the following document types: article, review and proceedings paper. These are the types of documents that are commonly cited, and feature in the calculation of the official IF in Journal Citation Reports (JCR). For simplicity, we will refer to these "citable" documents as "articles." We performed all of the analysis for citations received in 2010. Our results do not depend on the choice of year. For the analysis we selected 15,906 journals that have published 25 or more articles during the publication window (years 2008 and 2009). The cut was chosen to ensure well-sampled citation distributions, but the results are insensitive to the exact choice of the threshold. The total number of articles published in selected journals from 2008/09 is 2,352,554. The IF values computed from our data are smaller than those officially published by JCR by about 4%, because the latter includes citations to document types other than the articles (e.g., to editorials), as well as unmatched citations (citations for which the cited item is not identified other than that it belongs to that journal). See (Bar-Ilan (2010); Larivière et al. (2016); McVeigh and Mann (2009)) for details. In our analysis, we have adjusted the computed IF values by multiplying by a factor of 1.04. Accurate reproduction of the official IFs is not essential for our analysis because the missing citations are not expected to change the citation distribution (Larivière et al., 2016).

Our goal is to quantitatively compare some journal with another journal. We call the first journal the target journal (*t*), and the second journal the reference journal (*r*). The probability that a randomly drawn article from *t* will have a greater number of citations than an article drawn from *r*, i.e., the citation success index[1] ($S_{tr}$), will be:

$$S_{tr} = \sum_{c=0}^{\infty} [P_t(>c) + 1/2 P_t(c)] \, P_r(c) \quad (1)$$

where $P_t(>c)$ is the fraction of papers in journal *t* that have a number of citations larger than *c*, and $P_r(c)$ is the fraction of papers published in journal *r* that have received exactly *c* citations. Ties ($P_t(c) = P_r(c)$) are counted half of the times to retain symmetry ($S_{rt} = 1 - S_{tr}$). This probability is equivalent to the fraction of all "wins" if every article from target journal had been compared with every article from the reference journal. The calculation of the citation success index is carried out using the rank sum.

If two journals have identical citation distributions (the same fraction of articles having 0, 1, 2, etc. citations), the probability that an article drawn from the target journal will have more citations than an article from the reference journal will be exactly 50% (*S* = 0.5). In other words, the citation success index of 50% means that both journals do equally well. The citation success index of 90% (*S* = 0.9) suggests substantially better citation capacity of *t* over *r*.

---

[1] We express citation success index as a number (from 0-1) in equations, and as a percentage (from 0 to 100%) in text and tables.

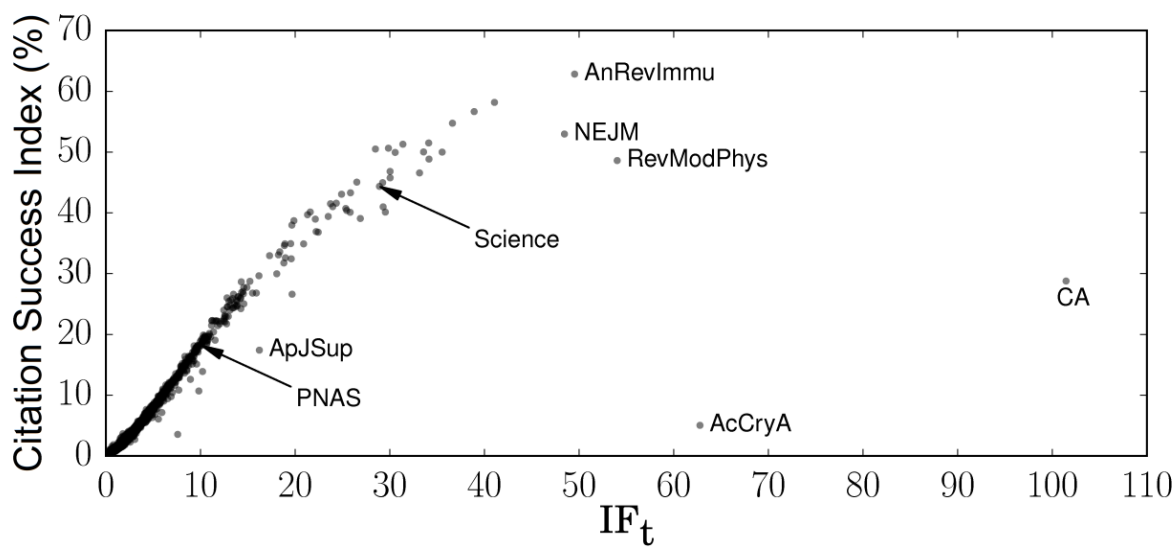

**Fig. 1.** *Citation success indices of 15,906 journals with respect to Nature.* Citation success index is the probability that a random article published in a target journal with impact factor $IF_t$, has more citations than a random article published in a reference journal, in this case *Nature*. A citation success index of 50% means that the two journals are the same in terms of their citation capacity. A citation success index of >50% means that the target journal does better than the reference journal. In the case of *Nature* there are few such journals. The relation is relatively narrow and the citation success index increases gradually as the $IF_t$ increases. Significant outliers and notable journals are designated by abbreviations and are discussed in the text.

## 3. Results

### 3.1. Citation success index and its relation to the impact factor

Fig. 1 shows the citation success index of different target journals with respect to *Nature*, which represents the reference journal in this case. Citation success index is plotted against the IF value of the target journal. The IF values of the various target journals range from 0 to 110, whereas the term of comparison, i.e., *Nature*, has an IF value of 35.5. If for some target journal the citation success index equals 50%, the articles in that journal are doing equally well in terms of citation capacity as the articles in *Nature*. Given that *Nature* is already one of the journals with the highest IF value (it is in the top 10 most highly ranked journals), it is not surprising that few journals have an advantageous citation success index over *Nature* ($S > 50\%$). Seven of the ten such journals are review journals (*Annual Review of Immunology, Nature Reviews: Molecular Cell Biology, Nature Reviews: Cancer, Nature Reviews: Immunology, Nature Reviews: Neuroscience, Physiological Reviews, Annual Review of Neuroscience*). The remaining three are *Cell, Nature Genetics,* and *New England Journal of Medicine*. Two journals with the highest IF values (*CA: A Cancer Journal for Clinicians* and *Acta Crystallographica Section A (AcCryA)*) actually have a citation success indices smaller than 50%, especially the latter (5%), for reasons that will be discussed shortly.

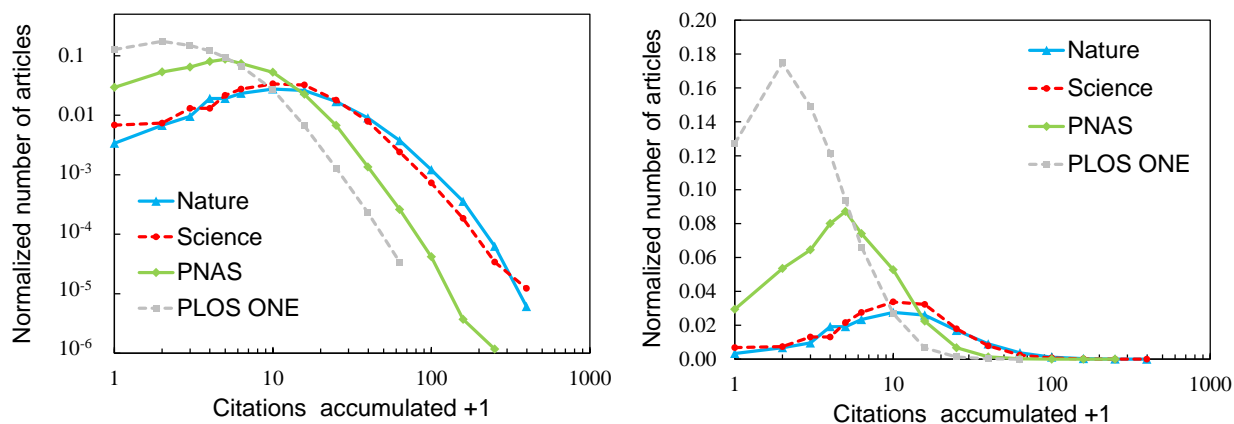

**Fig. 2.** *Citation distributions for four multidisciplinary journals (citations to papers published in 2008 and 2009 received in 2010).* The panels show the same distributions, but with a logarithmic (left) and linear (right) y-axes. Despite very different IF values (from IF = 4.5 for *PLOS ONE* to IF = 35.5 for *Nature*), the distributions have a significant overlap. They principally differ by the position of the peak and to some extent in the slope of the power-law tail of highly cited articles. The broadness of the distribution is the principal reason why the citation success index is not a very steep function of the IF ratio of journals being compared. Logarithmic binning is used to smooth the tails of distributions, following Milojević (2010).

    The remarkable feature of Fig. 1 is that the dispersion of the empirical points is relatively small: for a given IF value, the citation success index is narrowly distributed. There clearly exists a relation between the index (which depends on the shape of the citation distribution), and the IF. This is despite the fact that the IF is often perceived as an inadequate, or at least very limited, characterization of a journal's citation capacity, as discussed in Sec. 1. The reason why a tight relation exists nevertheless, lies in the empirical fact that *different journals with the same IF actually have quite similar citation distributions*. Significant exceptions exist, but they are rare. The reason why some journals in Fig. 1 scatter below what is otherwise a tight relation lies in the fact that their citation distributions are atypical compared to other journals of the same IF, usually because of a small number of very highly cited articles that boost the IF value. *Acta Crystallographica Section A (AcCryA)* is a unique and most drastic example of this effect. It owes its very high IF in 2010 (and 2009) to a single article published in 2008, which received 6890 citations in 2010 alone (it has a total of over 50,000 citations today). This paper is a review of SHELX software widely used in crystallography since 1976, which before the publication of this article did not have a citable reference (Schwarzenbach, Kostorz, McMahonc, & Strickland, 2010). The next most cited article in this journal have had only 21 citations in 2010. After this temporary surge, the IF of *AcCryA* has returned to a value around 2. The citation success index is not sensitive to such cases, and gives *AcCryA* an index over *Nature* of only 5%.

    We note that the relationship between the citation success index and the IF of the target journal is gradual. For example, Fig. 1 shows that an article published in a journal with an IF value equal to 20 (almost two times smaller than the IF of *Nature*), still has a 35% probability of receiving more citations than a random article published in *Nature*. A target journal with an IF value approximately equal to 10 (e.g., that of *PNAS*), leads to a citation success index of 17% over *Nature*. As the IF of the target journal decreases, so does the index. In more extreme cases, for example, for a journal with IF ~4 (e.g., *PLOS ONE*), the index is quite small (7%), but it is nevertheless not zero.

The reason why the relation between the IF of the target journal and the citation success index is not very steep lies in the fact that the citation distributions of journals tend to be broad and tend to overlap (Larivière et al., 2016; Redner, 1998; Stringer, Sales-Pardo, & Amaral, 2008). This can be appreciated from Fig. 2, where we show citation distributions of articles published in four major multidisciplinary journals: *Nature*, *Science*, *PNAS*, and *PLOS ONE*. The four journals have a wide range of IF values: from 35.5 for *Nature* to 4.5 for *PLOS ONE* (complete information is given in Table 1). Nevertheless, their citation distributions overlap to a large extent. *Nature, Science*, and *PNAS* have papers with anywhere between 0 and ~1,000 citations, while this range is between 0 and 200 for *PLOS ONE*. The relation between the IF values and the citation success index would have been steeper if the citation distributions were narrower. For example, if papers in *PLOS ONE* only had between 0 and 10 citations (which could still produce the actual IF = 4.5), while all papers in *Nature* had more than 10 citations (which could still result in IF = 35.5), then there would have been a null probability for any *PLOS ONE* paper to have more citations than any *Nature* paper. We also note that *Nature* and *Science* actually have very similar citation distributions, but the reason why *Science* has a somewhat smaller IF value than *Nature* (28.9 vs. 35.5) is due to the slightly smaller fraction of very highly cited papers than in *Nature*.

**Table 1**

Citation success indices among the pairs of four main multidisciplinary journals

| Journal | Nature | Science | PNAS | PLOS ONE | IF (from data) | IF (JCR 2010) | IF (JCR 2014) |
|---|---|---|---|---|---|---|---|
| **Nature** | - | 56% | 82% | 93% | 35.5 | 36.1 | 41.5 |
| **Science** | 44% | - | 79% | 92% | 28.9 | 31.4 | 33.6 |
| **PNAS** | 18% | 21% | - | 74% | 10.1 | 9.8 | 9.7 |
| **PLOS ONE** | 7% | 8% | 26% | - | 4.46 | 4.41 | 3.23 |

A generic entry of the table shows the citation success index of a journal listed in first column with respect to the journal in columns 2-5. The three rightmost columns of the table report: the 2010 IF derived from our bibliographic dataset, and the official IFs for years 2010 and 2014 published in the Journal Citation Report (JCR)

So far, we have discussed the relation between various journals and *Nature*. In Table 1, we present cross comparisons among four multidisciplinary journals. As expected, the biggest contrast is between *PLOS ONE* and *Nature*, in the sense that a random paper from *Nature* has 93% probability of having more citations than a random paper from *PLOS ONE*. Minimal difference is present between *Nature* and *Science*, with the citation success index of 56% in favor of *Nature*. We note that our calculations are based on 2010 data. The most recent IF values are slightly different: *Science* and especially *Nature* have higher IF values than they had in 2010, while *PNAS* is nearly the same, and *PLOS ONE* is lower. These changes will likely be reflected in somewhat more advantageous citation success indices of the first two journals with respect to the other two.

We also present a case study of citation success indices for publishing in biochemistry. The list of all journals in the JCR category Biochemistry & Molecular Biology was presented to an expert in the field who selected a comprehensive set of journals that were most relevant to his research field. From the list we selected a subset of 24 journals for which the citation success indices were calculated (see Appendix). These journals have IF values in the range 1.3 to 14.9. For journals in the intermediate impact range (IF ~5) the change in an IF of 1 (from 5 to 6) is associated with a marginal advantage of a higher-IF journal (citation success index of 55%).

*3.2. Universal relation between citation success index and impact factor*

The exact computation of the citation success index for a pair of journals requires the knowledge of full citation distributions of both journals. This is a clear limitation for wider implementation of this metric. Fortunately, as Fig. 1 shows, the citation success index and the IF values are related by a narrow function. This empirical fact permits the possibility of estimating the index quite precisely using only the IF values of the journals, which are readily available.

In Fig. 3, we show the citation success indices for four reference journals (*Science, PNAS, PLOS ONE,* and *Proceedings of the Royal Society A (PRSA)*), chosen to exhibit a fairly wide range of IF values, from 28.9 (*Science*) to 1.7 (*PRSA*). We now plot the index as a function of the *logarithm of the ratio* of the IF values of the target to the reference journal. The shape of the relation for all four reference journals is similar and has a characteristic sigmoid shape. When the IF ratio is high, the index approaches 100%. When the IF ratio is 1, the index is around 50%, as expected.

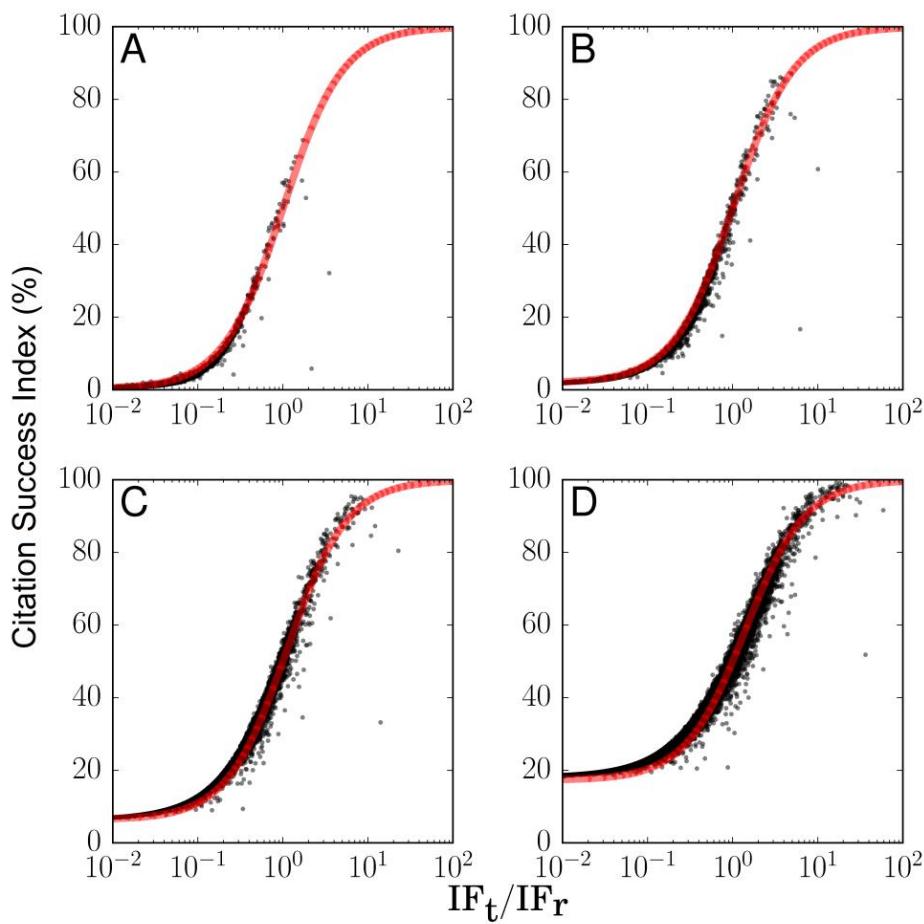

**Fig. 3.** *Citation success index as a function of impact factor ratios of the two journals.* Each panel shows a different reference journal (A = *Science;* B = *PNAS;* C = *PLOS ONE;* D = *PRSA*), spanning a large range of impact factors. Trends are similar except for the lower plateau that depends on the fraction of uncited articles in the reference journal (see text). All trends can be well described by a modified logistic function (Eq. 3, red curves) with a variable starting point (the plateau) and the exponent that is approximately independent of the reference journal.

The main difference between the four curves is in the location of the lower asymptote – the probability that an article from a target journal with a very small IF will have more citations than a paper from the reference journal. This plateau probability is close to zero for a high-IF journal like *Science*, but becomes as high as 20% for *PRSA* (IF = 1.7). The non-zero plateau is due to the uncited papers in the reference journals. *PLOS ONE* and *PRSA* publish a non-negligible proportion of papers that do not receive citations (at least in the time window used for calculating the IF), so that even a target journal with IF = 0 (no paper having received any citation) will be tied with uncited articles from the reference journal. Because the ties count as "greater than" half of the time, the plateau will be located at ½ of the "uncited fraction" ($f_0$) of the reference journal.

The existence of a plateau that depends on the uncited fraction seems to prevent the construction of a citation success index function that would only depend on easily available IFs. Fortunately, Fig. 4A shows that the fraction of uncited articles is itself a tight function of the IF, a feature noted in some previous studies (Moed, Van Leeuwen, & Reedijk, 1999; Schubert & Glänzel, 1983; Weale, Bailey, & Lear, 2004). This is another consequence of the fact that the journals with the same IFs have largely similar citation distributions. For journals with IF ~1, the uncited fraction is around 50%. The tightness of the scatter plot suggests that a suitable functional form could allow a relatively precise determination of $f_0$ from the IF alone. We find that $f_0$ is described remarkably well by the generalized logistic function (to be accurate, the function is logistic when the IF is expressed as a logarithm):

$$f_0 = \frac{1}{(1 + qIF^\alpha)^\beta} \qquad (2)$$

where the values of the factor $q$ and exponents $\alpha$ and $\beta$ are: $\alpha = 0.94$, $\beta = 2.37$, and $q = 0.33$.

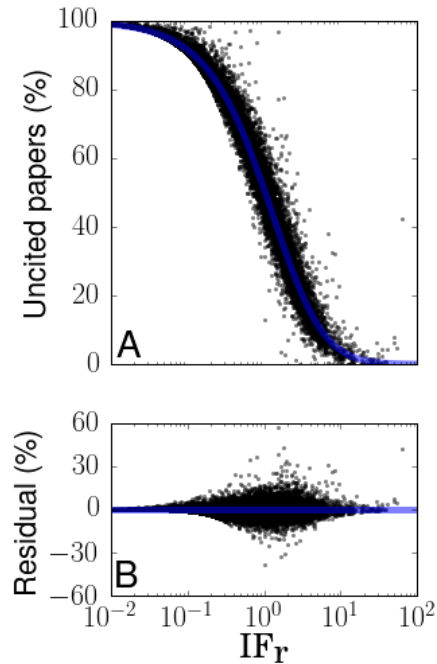

**Fig. 4.** *Relation between the fraction of uncited articles in a journal and its impact factor.* (A) For every journal, we computed the fraction $f_0$ of papers that have accumulated zero citations one or two years after their publication, and plotted it as function of the IF value of the journal. The blue curve is given by the generalized logistic function of Eq. 2. (B) Residuals are symmetrically distributed around the fit (blue curve and line), and their value is independent of the IF value of the journal.

At this point we have all the ingredients necessary to establish a relation between the citation success index and the IF ratio. This is given by a logistic function with a positive lower asymptote:

$$S = f_0/2 + \frac{1 - f_0/2}{1 + qx^{-k}} \qquad (3)$$

where $x = IF_t/IF_r$ is the ratio of IFs of target and reference journals, and $f_0$ is the uncited rate of the reference journal that can be evaluated from Eq. 2, or read off from Fig. 4. Factor $q$ is required to ensure that $S = 0.5$ when $x = 1$, and equals $q = 1/(1 - f_0)$.

Fitting Eq. 3 to the data in order to determine $k$ is performed as follows. Citation success indices are averaged in equal bins in log $x$ of 0.05. Binning ensures that equal weight is given to the journals with different IF ratios. Fitting is performed by minimizing the square deviations of indices with respect to the fitting function. The fitting has only one parameter, the exponent $k$. Fig. 5 shows that $k$ only weakly depends on the reference journal, giving Eq. 2 a *universal* character. On average, it takes the value $k = 1.23$. The citation success index is therefore a function of two independent variables $IF_r$ and $x$. When the uncited rate of the reference journal is low ($IF_r \geq 10$) or when $x \geq 1$, Eq. 3 simplifies to:

$$S = \frac{1}{1 + x^{-k}} \qquad (4)$$

i.e., the citation success index then depends solely on the IF ratio $x$.

In Fig. 6 we show the citation success index matrix for journals with IF > 3 and the residuals with respect to true values when the indices are obtained using only Eq. 2 and 3, with the value of $k$ fixed to 1.23. The residuals are small (a couple of percent) and symmetric.

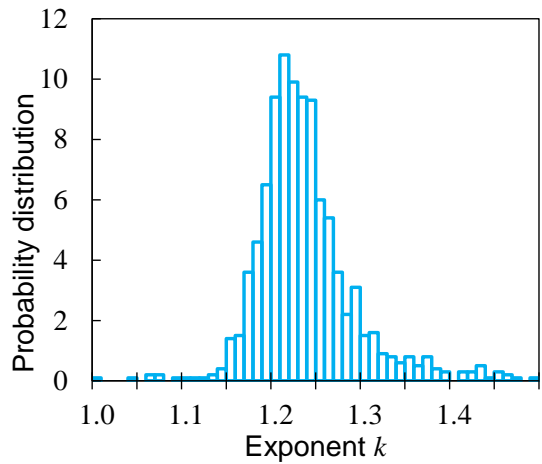

**Fig. 5.** *The distribution of the best-fitting values of exponent k in Eq. 3, describing the shape of the function between the citation success index and impact factor ratio*. Based on 1,400 journals with IF > 3. The exponents take a relatively small range of values attesting to the universality of the citation success index – IF ratio relation shown in Fig. 3.

To facilitate the calculation of citation success indices, we provide a web calculator (http://tinyurl.com/hxgnz4f), which only requires a user to input the IFs of two journals.

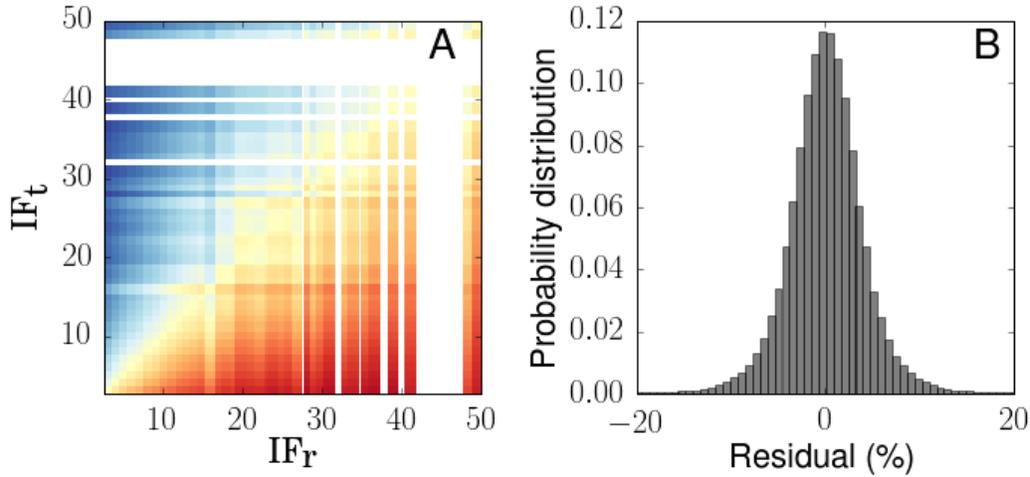

**Fig. 6.** *The citation success index matrix for journals with IF > 3.* In panel A, we show the citation success index of a target journal with an IF value equal to $IF_t$, compared to a reference journal with an IF value equal to $IF_r$. To generate this figure, we consider only the 1,400 journals with IF> 3. Colors range from red (index = 0%) to blue (index = 100%). The empirical values in panel A are very well reproduced by our Eqs. 2 and 3, which only depend on the IFs. The residuals are typically small (a couple of percent), and are distributed symmetrically around zero (panel B).

### 4. Discussion and conclusions

Recently, Larivière et al. (2016) have pointed out that IF alone hides important information regarding the broadness of citation distributions, and have shown that even journals with quite different IFs have overlapping citation distributions. Our measure, the citation success index, addresses this concern and allows for an intuitive comparison of citation distributions of pairs of journals. It provides an answer to a question: what is the probability that an article from one journal has more citations than the article from another journal? This measure depends on the shape of citation distributions, but being a single number is more practical as a metric than the comparison of detailed citation distributions.

We have shown that the relation between the citation success index and the IF ratio of the journals being compared is relatively tight, a consequence of the fact that citation distributions of journals with the same IFs are quite similar (Radicchi, Fortunato, & Castellano, 2008; Stringer et al., 2008). For the same reason, the fraction of articles with zero citations can be predicted from the IF alone. Atypical distributions are rare, which is why there are few outliers in the citation success index – IF ratio relation.

Furthermore, we have shown that the citation success index – IF ratio relation is a universal function of the IFs or the target and reference journals. When the IF ratio > 1 the citation success index largely depends only on the ratio. For example, an article from journal A will have ~70% probability of having more citations than an article from journal B regardless of whether the IF of A is 10 and of B is 5, or if A is 30 and B is 15. Essentially, we demonstrate that the *relative* differences in IFs are more relevant in characterizing the citation capacity of journals than their absolute differences.

The citation success index – IF ratio relation grows gradually. Thus, the citation success index is not substantially different for small relative differences in IF (IF ratios <2). The fundamental reason for the gradual change lies in the fact that even journals with very different IFs have broad and largely overlapping citation distributions (Fig. 2). There has been a vocal debate as to whether IFs are pertinent in the context of impacts of individual articles published therein. Our citation success index measure allows us to provide a nuanced answer to this question: 'definitely yes', when the IFs differ by more than a factor of several, and 'not much' when the differences are within the factor of two. This is an important lesson for the authors who strive to publish their work in journals with moderately higher impact factors than alternative venues.

**Acknowledgments**

The work uses Web of Science data by Thomson Reuters provided by the Network Science Institute and the Cyberinfrastructure for Network Science Center at Indiana University. We thank Andras Muhlrad for selecting the biochemistry journals and William Coates for copyediting. FR acknowledges NSF grant SMA-1446078.

**Appendix**

**Table 2.**

Citation success index matrix for biochemistry journals. The index of an article in a biochemistry journal in row *m* (numbered 1-24) with respect to the journal in column *n* (numbered 1-24). For example, an article published in *Biomacromolecules* (row 12) has a 66% probability of having more citations than an article in *Biochemistry* (column 20).

| | Journal | IF (from data) | IF (JCR2010) | 1 | 2 | 3 | 4 | 5 | 6 | 7 | 8 | 9 | 10 | 11 | 12 | 13 | 14 | 15 | 16 | 17 | 18 | 19 | 20 | 21 | 22 | 23 | 24 |
|---|---|---|---|---|---|---|---|---|---|---|---|---|---|---|---|---|---|---|---|---|---|---|---|---|---|---|---|
| 1 | NATURE CHEMICAL BIOLOGY | 14.9 | 15.8 | 50 | 52 | 65 | 63 | 62 | 66 | 77 | 78 | 79 | 81 | 80 | 79 | 81 | 80 | 80 | 83 | 82 | 85 | 86 | 89 | 89 | 89 | 92 | 95 |
| 2 | NATURE STRUCTURAL & MOLECULAR BIOLOGY | 13.8 | 13.7 | 48 | 50 | 65 | 62 | 61 | 64 | 77 | 78 | 79 | 82 | 80 | 79 | 81 | 81 | 80 | 85 | 83 | 86 | 88 | 90 | 90 | 91 | 94 | 97 |
| 3 | CRITICAL REVIEWS IN BIOCHEMISTRY AND MOLECULAR BIOLOGY | 10.8 | 10.1 | 35 | 35 | 50 | 48 | 45 | 50 | 61 | 62 | 63 | 67 | 68 | 64 | 65 | 67 | 68 | 72 | 71 | 72 | 77 | 78 | 80 | 81 | 87 | 93 |
| 4 | TRENDS IN BIOCHEMICAL SCIENCES | 10.5 | 10.4 | 37 | 38 | 52 | 50 | 48 | 52 | 63 | 64 | 66 | 69 | 69 | 66 | 68 | 69 | 69 | 73 | 72 | 74 | 78 | 79 | 80 | 81 | 86 | 91 |
| 5 | EMBO JOURNAL | 10.3 | 10.1 | 38 | 39 | 55 | 52 | 50 | 54 | 66 | 68 | 69 | 72 | 72 | 70 | 71 | 72 | 72 | 77 | 75 | 78 | 81 | 82 | 84 | 85 | 89 | 94 |
| 6 | CURRENT OPINION IN CHEMICAL BIOLOGY | 9.4 | 9.3 | 34 | 36 | 50 | 48 | 46 | 50 | 61 | 62 | 64 | 67 | 67 | 64 | 66 | 67 | 67 | 71 | 70 | 72 | 76 | 77 | 78 | 79 | 84 | 90 |
| 7 | MOLECULAR AND CELLULAR BIOLOGY | 6.3 | 6.2 | 23 | 23 | 39 | 37 | 34 | 39 | 50 | 51 | 53 | 57 | 58 | 54 | 55 | 57 | 58 | 63 | 62 | 63 | 70 | 70 | 72 | 74 | 80 | 88 |
| 8 | FASEB JOURNAL | 6.1 | 6.5 | 22 | 22 | 38 | 36 | 32 | 38 | 49 | 50 | 52 | 56 | 57 | 52 | 54 | 56 | 57 | 62 | 60 | 62 | 68 | 68 | 71 | 73 | 79 | 87 |
| 9 | CHEMISTRY & BIOLOGY | 5.8 | 5.8 | 21 | 21 | 37 | 34 | 31 | 36 | 47 | 48 | 50 | 54 | 55 | 50 | 52 | 54 | 55 | 59 | 58 | 59 | 66 | 66 | 68 | 70 | 77 | 85 |
| 10 | MOLECULAR BIOLOGY AND EVOLUTION | 5.7 | 5.5 | 19 | 18 | 33 | 31 | 28 | 33 | 43 | 44 | 46 | 50 | 52 | 46 | 48 | 50 | 51 | 56 | 55 | 55 | 63 | 62 | 65 | 67 | 74 | 83 |
| 11 | ACS CHEMICAL BIOLOGY | 5.7 | 5.7 | 20 | 20 | 32 | 31 | 28 | 33 | 42 | 43 | 45 | 48 | 50 | 45 | 46 | 49 | 49 | 53 | 53 | 53 | 60 | 59 | 62 | 64 | 71 | 79 |
| 12 | BIOMACROMOLECULES | 5.6 | 5.3 | 21 | 21 | 36 | 34 | 30 | 36 | 46 | 48 | 50 | 54 | 55 | 50 | 52 | 54 | 55 | 60 | 59 | 60 | 67 | 66 | 69 | 71 | 78 | 86 |
| 13 | JOURNAL OF BIOLOGICAL CHEMISTRY | 5.5 | 5.3 | 19 | 19 | 35 | 32 | 29 | 34 | 45 | 46 | 48 | 52 | 54 | 48 | 50 | 52 | 54 | 58 | 57 | 58 | 66 | 65 | 68 | 70 | 77 | 85 |
| 14 | BIOCHIMICA ET BIOPHYSICA ACTA-BIOENERGETICS | 5.4 | 5.1 | 20 | 19 | 33 | 31 | 28 | 33 | 43 | 44 | 46 | 50 | 51 | 46 | 48 | 50 | 51 | 55 | 54 | 55 | 62 | 61 | 64 | 66 | 73 | 82 |
| 15 | BIOCHIMICA ET BIOPHYSICA ACTA-MOLECULAR AND CELL BIOLOGY OF LIPIDS | 5.4 | 5.1 | 20 | 20 | 32 | 31 | 28 | 33 | 42 | 43 | 45 | 49 | 51 | 45 | 46 | 49 | 50 | 54 | 54 | 54 | 61 | 60 | 63 | 66 | 73 | 82 |
| 16 | BIOCHEMICAL JOURNAL | 5.1 | 5.0 | 17 | 15 | 28 | 27 | 23 | 29 | 37 | 38 | 41 | 44 | 47 | 40 | 42 | 45 | 46 | 50 | 50 | 49 | 58 | 56 | 60 | 63 | 70 | 80 |
| 17 | BIOCHIMICA ET BIOPHYSICA ACTA-MOLECULAR CELL RESEARCH | 5.0 | 4.7 | 18 | 17 | 29 | 28 | 25 | 30 | 38 | 40 | 42 | 45 | 47 | 41 | 43 | 46 | 46 | 50 | 50 | 50 | 58 | 56 | 59 | 62 | 69 | 78 |
| 18 | JOURNAL OF MOLECULAR BIOLOGY | 4.2 | 4.0 | 15 | 14 | 28 | 26 | 22 | 28 | 37 | 38 | 41 | 45 | 47 | 40 | 42 | 45 | 46 | 51 | 50 | 50 | 59 | 57 | 61 | 64 | 71 | 81 |
| 19 | BIOCHIMIE | 3.9 | 3.8 | 14 | 12 | 23 | 22 | 19 | 24 | 30 | 32 | 34 | 37 | 40 | 33 | 34 | 38 | 39 | 42 | 42 | 41 | 50 | 47 | 51 | 54 | 61 | 72 |
| 20 | BIOCHEMISTRY | 3.3 | 3.2 | 11 | 9.9 | 22 | 21 | 18 | 23 | 30 | 32 | 34 | 38 | 41 | 34 | 35 | 39 | 40 | 44 | 44 | 43 | 53 | 50 | 54 | 57 | 65 | 76 |
| 21 | FEBS JOURNAL | 3.2 | 3.1 | 11 | 9.6 | 20 | 20 | 16 | 22 | 28 | 29 | 32 | 35 | 38 | 31 | 32 | 36 | 37 | 40 | 41 | 39 | 49 | 46 | 50 | 53 | 61 | 72 |
| 22 | ARCHIVES OF BIOCHEMISTRY AND BIOPHYSICS | 3.1 | 3.0 | 11 | 9.5 | 19 | 19 | 15 | 21 | 26 | 27 | 30 | 33 | 36 | 29 | 30 | 34 | 34 | 37 | 38 | 36 | 46 | 43 | 47 | 50 | 57 | 69 |
| 23 | JOURNAL OF BIOCHEMISTRY | 2.2 | 2.1 | 7.7 | 6 | 13 | 14 | 11 | 16 | 20 | 21 | 23 | 26 | 29 | 22 | 23 | 27 | 27 | 30 | 31 | 29 | 39 | 35 | 39 | 43 | 50 | 61 |
| 24 | BIOCHEMISTRY-MOSCOW | 1.3 | 1.4 | 4.7 | 3 | 7 | 9 | 6 | 10 | 12 | 13 | 15 | 17 | 21 | 14 | 15 | 18 | 18 | 20 | 22 | 19 | 28 | 24 | 28 | 31 | 39 | 50 |